\documentclass{raa}            

\usepackage{graphicx,times}             
\usepackage[authoryear]{natbib}

\usepackage{amssymb,amsmath}
\bibpunct{(}{)}{;}{a}{}{,}

\usepackage[pagebackref=true]{hyperref}

\begin{document}

  \title{A Study on Non-coplanar Baseline Effects for Mingantu Spectral Radioheliograph}

   \volnopage{Vol.0 (20xx) No.0, 000--000}      
   \setcounter{page}{1}          

   \author{Qiu-ping Yang 
      \inst{1,2,5}
   \and Feng Wang
      \inst{3,1,5}
   \and Hui Deng
      \inst{3}
   \and Ying Mei
	\inst{3}
   \and Wei Wang
	\inst{4}
   }

   \institute{Yunnan Observatories, Chinese Academy of Sciences,
Kunming 650216, P.R. China; {\it fengwang@gzhu.edu.cn}\\
        \and Faculty of Information Engineering and Automation, Kunming University of Science and Technology,
No.727 South Jingming Rd., Chenggong District, Kunming 650500, China
             \\
	\and Center of Astrophysics, Guangzhou University,
No.230, Waihuanxi Rd, Panyu District, Guangzhou 510006, China
		\\
	\and National Astronomical Observatory, Chinese Academy of Sciences 
20A Datun Rd., Chaoyang District, Beijing 100012, China
		\\
\and University of Chinese Academy of Sciences,
19A Yuquan Rd., Shijingshan District, Beijing 100049, China
             \\		
\vs\no
   {\small Received 20xx month day; accepted 20xx month day}}

\abstract{ As a dedicated solar radioheliograph, the MingantU SpEctral RadioHeliograph (MUSER) has a maximum baseline of more than 3000 meters and a frequency range of 400 MHz -- 15 GHz. According to the classical radio interferometry theory, the non-coplanar baseline effect (i.e., w-term effect) would be considered and calibrated for such a radio instrument. However, little previous literature made the qualitative or quantitative analyses on w-term effects of solar radioheliograph in-depth. 
This study proposes a complete quantitative analysis of w-term effects for the MUSER. After a brief introduction of the MUSER, we systematically investigate the baseline variations over a year and analyze the corresponding variations of w-term. We further studied the effects of the w-term in the imaging for the specified extended source, i.e., the Sun. We discussed the possible effects of the w-term, such as image distortion and so on. The simulated results show that the w-term is an essential and unavoidable issue for solar radio imaging with high spatial resolution.
}

   \authorrunning{Q.-P. Yang, F. Wang, H. Deng, Y. Mei, \& W. Wang}            
   \titlerunning{A Study on Non-coplanar Baseline Effects for MingantU SpEctral RadioHeliograph}  

   \maketitle

%
%
\section{Introduction}           
\label{sect:intro}

Aperture-synthesis imaging technology~\citep{Ryle1962} has been widely used in radio astronomy because of its advantages, such as high resolution and quick imaging. 
Radioheliograph is a type of radio interferometer specially designed to observe the Sun. Several radioheliograph telescopes, such as Nobeyama Radioheliograph (NoRH)~\citep{Nakajima1994}, the Nancay Radioheliograph (NRH) \citep{Kerdraon1997,Kerdraon2011}, the Gauribidanur Radioheliograph~\citep{Ramesh1998}, the Siberian Solar Radio Telescope \citep{Grechnev2003} (SSRT\footnote{\url{http://en.iszf.irk.ru/The_Siberian_Solar_Radio_Telescope}}), and the Frequency-Agile Solar Radiotelescope(FASR,~\citep{Bastian2003, Gary2003} had been constructed or were under-constructed for monitoring solar activities.

The MUSER is a solar dedicated radio interferometric array with high temporal, spatial, and spectral resolutions. It can simultaneously perform the spectral and imaging observations of the full Sun in a wide frequency range\citep{Yan2004,Yan2015, Yan2021}. The MUSER has two sub-arrays: The MUSER-I contains 40 antennas with a 4.5m diameter for low frequency ranging from 400 MHz to 2 GHz. The MUSER-II contains 60 antennas with a 2m diameter operating frequency ranging from 2 GHz to 15 GHz. The specifications of the MUSER are listed in Table~\ref{basicparameters} \citep{Yan2011, Wang2015}. All the 100 antennas are located on three log-spiral arms with a maximum baseline length of about 3 km in north-south and east-west directions (longitude=115.2505$^{\circ}$, latitude=42.211833333$^{\circ}$, altitude=1365.0 meters).

\begin{table*}[htbp]
\caption{MUSER Characteristic and Performance} 
\label{basicparameters}
\centering
\begin{tabular}{ccc}
  \hline
  Parameters 			& MUSER-I            & MUSER-II\\
  \hline
  Antenna array            & 40 (4.5 meters)    &  60 (2 meters)\\
  Frequency range      &  0.4 GHz$-$2.0 GHz &  2.0 GHz$-$15 GHz\\
  Frequency resolution   &  64 channels   &  520 channels\\
  Channel bandwidth    & \multicolumn{2}{c}{25 MHz}\\
  Time resolution      & 25 ms & 206.25 ms\\
  Spatial resolution   & ${51.6}''- {10.3}''$ & ${10.3}''- {1.4}''$\\
  Maximum Baseline length      & \multicolumn{2}{c}{ more than 3000 m }   \\
  Polarization         & \multicolumn{2}{c}{Dual circular left and right}\\
  Dynamic range        & \multicolumn{2}{c}{25 db (snapshot)}\\
  Field of view        & \multicolumn{2}{c}{$0.6^{\circ} - 7^{\circ}$}\\
  \hline
\end{tabular}
\end{table*}

To obtain the high-quality image, a series of applications~\citep{Wang2015,Wei2016,Mei2017, Wang2019}, such as data preparation, calibration, gridding, and final image clean, had been developed for the MUSER. Meanwhile, many possible errors were also calibrated and corrected carefully. For example, the transport delay of the optical fiber and the phase tracking error of calibrated source (i.e., geostationary satellite) have been calibrated to guarantee the final quality image.

MUSER has a maximum baseline of more than 3000 meters and a frequency range of 400 MHz – 15 GHz. According to the classical radio interferometry theory, the non-coplanar baseline effect (i.e., w-term effect) would be considered and calibrated. However, there are few qualitative and quantitative analyses on the non-coplanar baseline effect (i.e., w-term effect) for solar radio observation. One possible reason is that the sun's apparent diameter from the earth is only about 32 arc minutes which is far from the concept of wide-field imaging. 

The w-term is a classical problem in radio interferometric imaging. There are several methods to deal with the w-term effects of imaging. \cite{Perley1989} proposed the three-dimensional Fourier transform. However, because of its high computational cost, this method is becoming more and more complicated when dealing with the massive amount of observing data. The polyhedron imaging method, or image plane faceting, was introduced by \cite{Perley1989}. An alternative faceting method is $uv$ plane faceting~\citep{Sault1999}. The w-projection algorithm is proposed based on the idea that $w$ could project onto the $w=0$ plane~\citep{Cornwell2005, Cornwell2008}. The A-projection algorithm~\
p{Bhatnagar2008} extended upon this, also including instrumental effects. W-stacking~\citep{Humphreys2011, Offringa2014} is another algorithm to deal with the w-term. 

The rest of the paper is organized as follows. In section 2, a detailed discussion of the w-term that arose from the non-coplanar baseline effects on MUSER and the focus problem of w-term effects quantitative analyses. The w-term effects on MUSER imaging performance are discussed in section 3. Section 4 concludes the study and presents specific aspects of our future work.

\section{non-coplanar baseline effects Of MUSER}
\subsection{w-term issue of MUSER}
For a radio interferometer, the received visibilities can be calculated by\citep{Thompson2008} 
\begin{equation}
\label{viscal}
\centering
V(u,v,w)=\int\int \frac{I(l,m)}{\sqrt{1-l^{2}-m^{2}}}e^{-2\pi i [ul+vm+w(\sqrt{1-l^{2}-m^{2}}-1)]}dldm
\end{equation}
where $(u,v,w)$ are the baseline coordinates, and the direction cosines $(l,m,n)$ denote a position on the celestial sphere defined by $l^2+m^2+n^2=1$. If the fields of view is sufficiently small, or the observation time is short enough (snapshot observations), it can be simplified by the approximation that $\sqrt{1-l^2-m^2}\approx 1$. Therefore, $2\pi w(\sqrt{1-l^2-m^2}-1)$ is much less than unity. In such a situation, all visibility data can be approximately considered to lie on a plane\citep{Bhatnagar1999}. Equation~\ref{viscal} could be simplified as

\begin{equation}
\centering
V(u,v,w)=\int \int I(l,m)e^{-2\pi i(ul+vm)}dldm
\label{viscal2d}
\end{equation}  

As observation fields of view becomes large, $2\pi w(\sqrt{1-l^2-m^2}-1)$ can no longer be ignored, thereby leading to w-term issues. \cite{Perley1999} proposed that w-term is the phase shift due to the dependence of direction cosines and the value of $w$, which may
be presented as  
\begin{equation}\label{wterm}
\centering
\phi(w)=2\pi w(\sqrt{1-{l}^2-{m}^2}-1)
\end{equation}

The maximum value of the term  $2\pi w(\sqrt{1-l^2-m^2}-1)$ could be expressed as $\frac{B\lambda}{D^2}$, where B is the longest baseline, $\lambda$ is the observing wavelength, and $D$ is the diameter of the elements of the array. The w-term can be ignored when
\begin{equation}\label{wless}
    \vert w(\sqrt{1-l^2-m^2}-1) \vert \ll 1
\end{equation}

\begin{figure*}[htbp]
    \centering
    \includegraphics[width=0.6\textwidth]{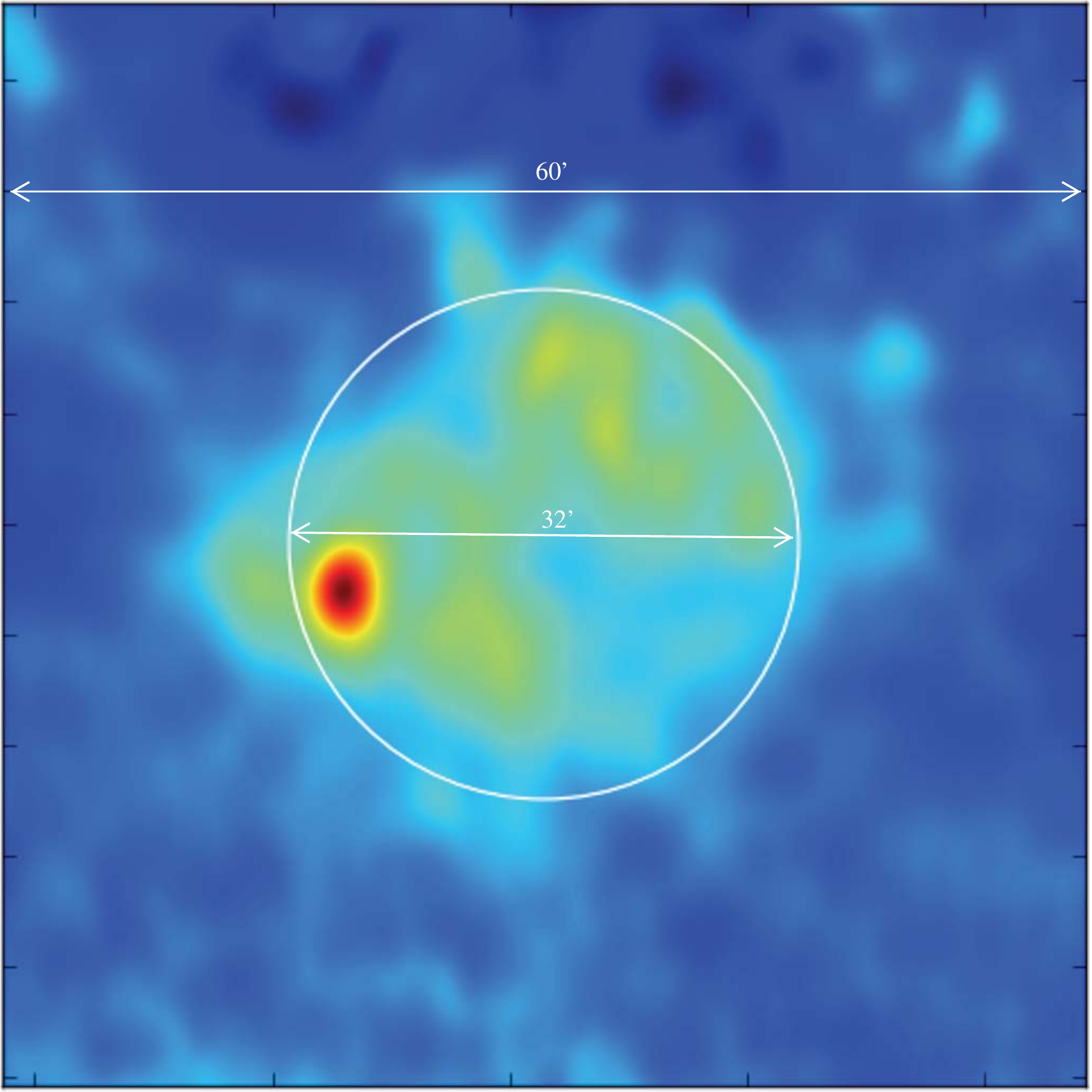}
    \caption{The observational clean image of MUSER-I, 1.7125GHz, 2015-11-01 04:08:49.354161240(UTC).}
    \label{fig_sun_image}
\end{figure*}

Unlike ordinary radio interferometric arrays, MUSER’s observation target is the Sun. Due to the revolution, the apparent position of the Sun changes continuously during the year, thus changing the value of $(u,v,w)$.
A schematic diagram of the MUSER imaging result is given in Figure~\ref{fig_sun_image}. The white circle range is the solar disk. Since the diameter of the Sun is about 32 arc minutes, the field of view of MUSER observation imaging is considered to be 1 square degree under normal conditions, which can effectively show the imaging results at different heights of the Sun.

According to the study of ~\cite{taylor1999synthesis}, we can allow the w-term effect to be as high as 0.1 radians without introducing severe errors in the image. When the field of view is 32 arc minutes, therefore, a simple calculation shows that the value of $w$ needs to be less than 1154.06 wavelength numbers for Equation~\ref{wless} to hold.

\begin{figure*}[htbp]
\centering
\includegraphics[scale=0.5, keepaspectratio=true]{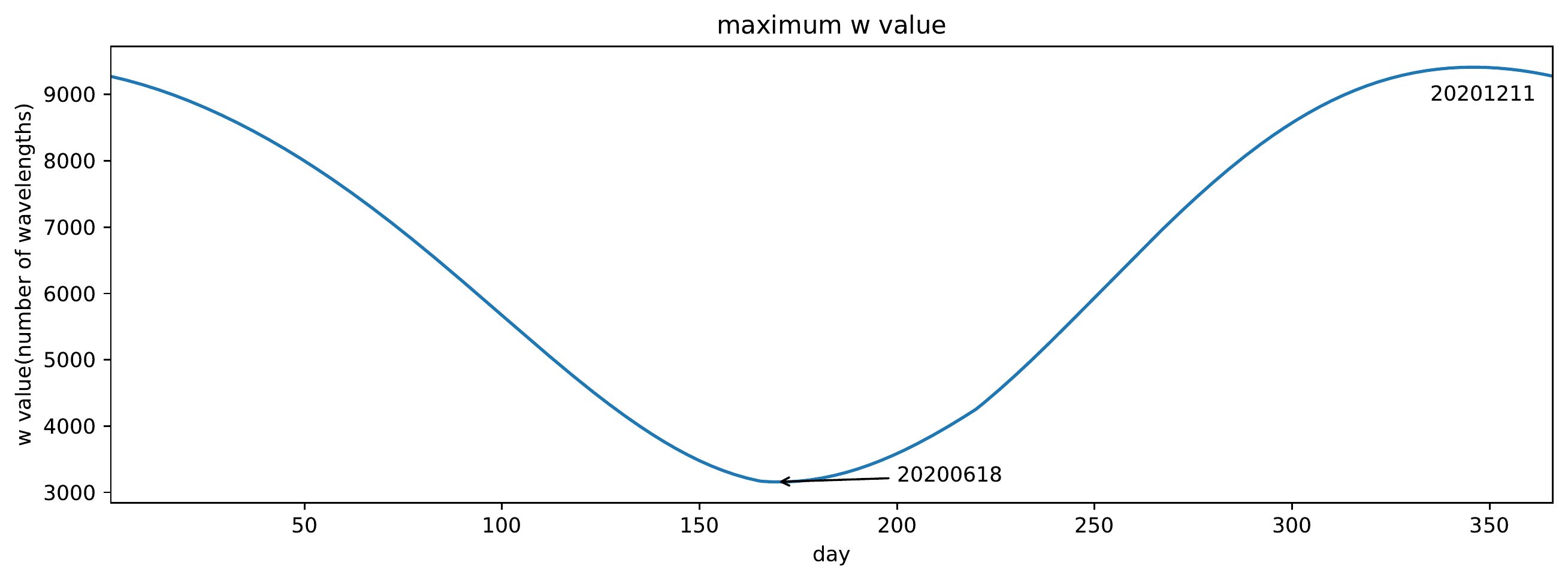}
\caption{The maximum w value of MUSER at 12:00 noon (CST) every day in the year 2020}
\label{fig_wmax2020}
\end{figure*}

To quantify the w-term effects of MUSER, we calculated the w-term variation for different frequencies of MUSER-I. Considering that the low frequencies are more susceptible to the w-term, we calculated the $w$-value variation at noon (China Standard Time (CST), UTC+8) for each day in a year(2020) using the 1 GHz observation of MUSER-I as an example.
Figure~\ref{fig_wmax2020} shows that the maximum value of $w$ in 2020 was about 9412.12 wavelength numbers, which occurred on December 11, and similarly, the minimum value was 3156.74  wavelength numbers on June 18. 

\begin{figure}[h]
  \begin{minipage}[t]{0.495\linewidth}
    \centering
    \includegraphics[width=60mm]{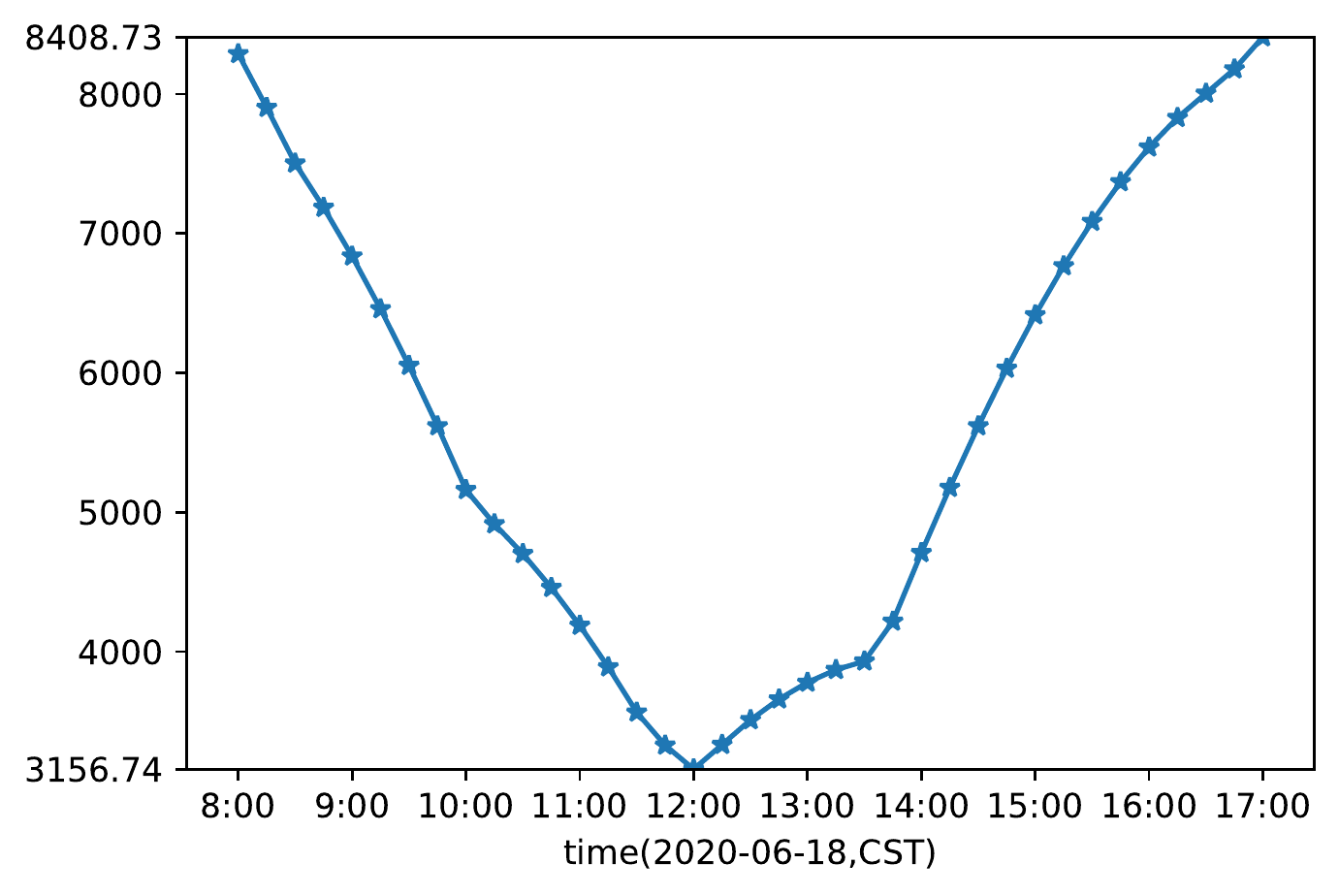}
  \end{minipage}
  \begin{minipage}[t]{0.495\linewidth}
    \centering
    \includegraphics[width=60mm]{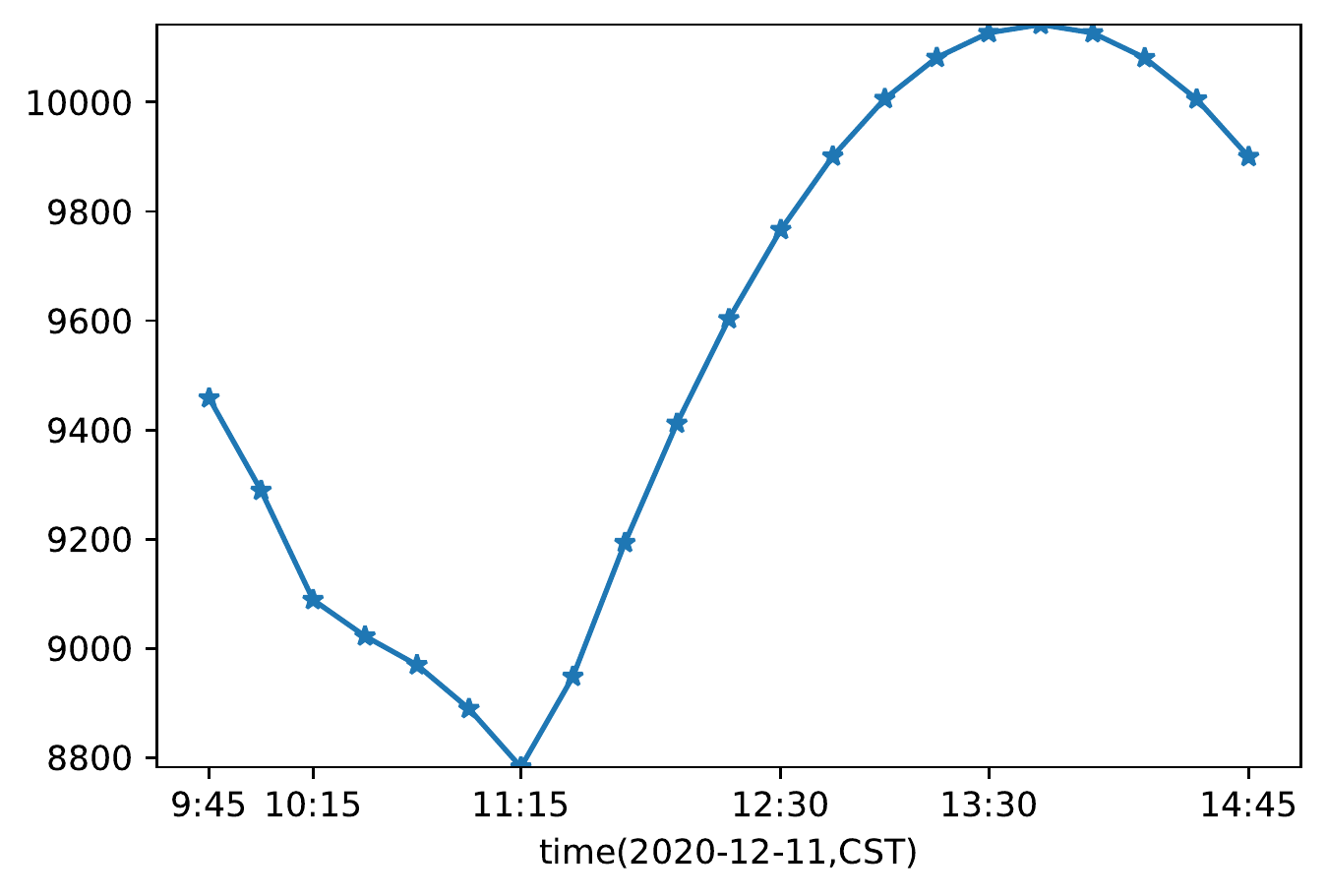}
  \end{minipage}
  \caption{ Maximum $w$ value from 8:00(am) to 17:00(pm)(CST), w unit:number of wavelength.}
  \label{fig_w_variation}
\end{figure}

We calculated the variation of the $w$ value during a day as well, giving the distribution of the maximum $w$ value from 8:00 to 17:00 CST time with a frequency of 1GHz(see
Figure~\ref{fig_w_variation}).
In practice, observations are generally considered to be made when the elevation angle of the Sun is greater than 15 degrees. Therefore, we can obtain that the observable time on December 11 was from 9:45 to 14:45(CST), with the $w$ peak value of 10,141.84 at 13:45 and the minimum value of 8,783.24 at 11:15; the $w$ peak value on June 18 was 8,408.73 at 17:00, and the minimum value was 3,156.74 at noon.

The results show that the $w$ value of MUSER was always greater than 1154.06 wavelength numbers during the year. Obviously, for the observation of the MUSER-I with 400 MHz, the value of w-term would be much larger than unity. W-term needs to be more explicitly dealt with, or the operation of the instrument will be significantly restricted.

\subsection{Quantitative analysis of Phase shift}

\begin{figure}[h]
    \begin{minipage}[]{0.495\linewidth}
    \centering
    \includegraphics[width=65mm]{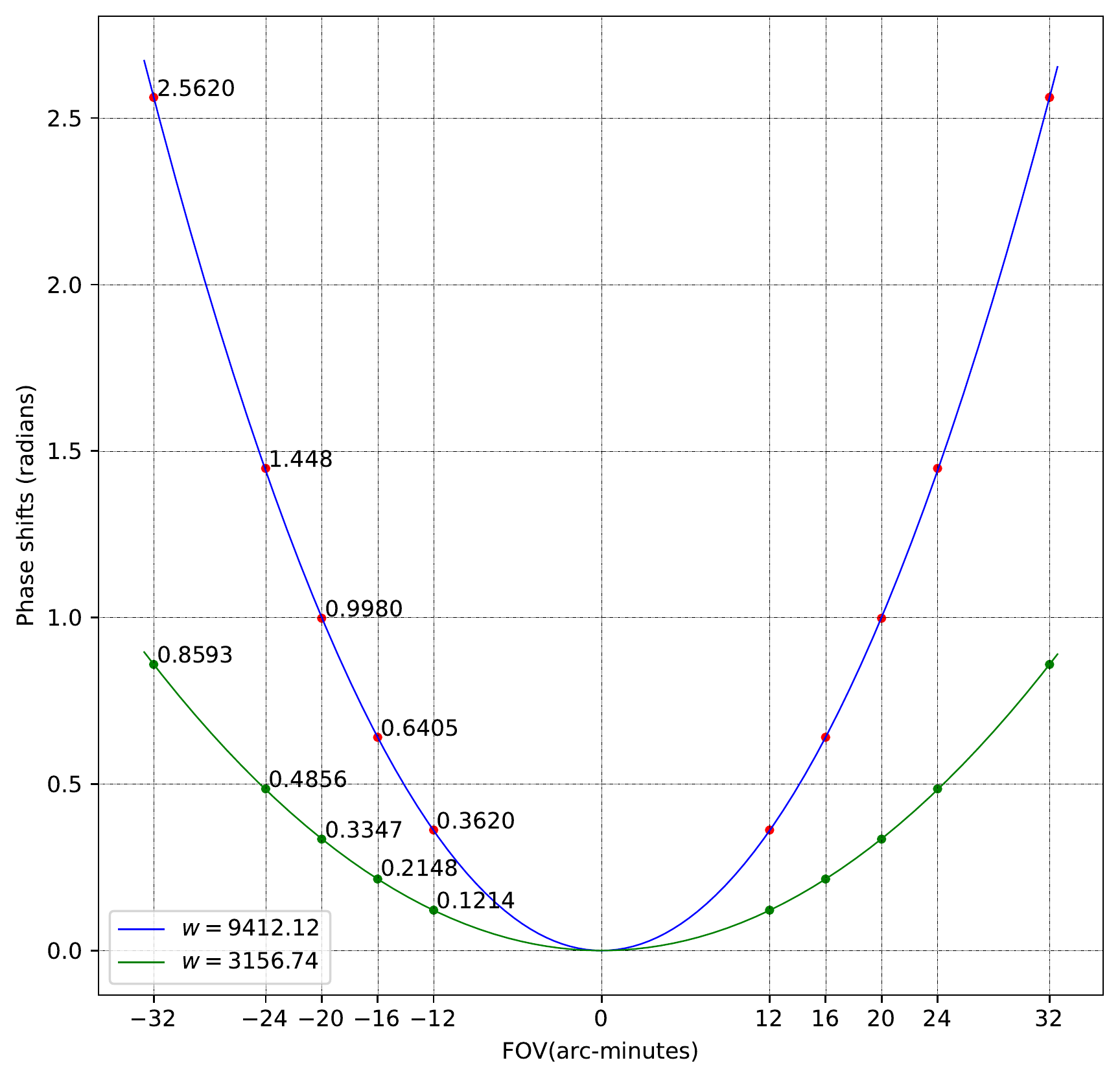}
  \end{minipage}
  \begin{minipage}[]{0.495\linewidth}
    \centering
    \includegraphics[width=80mm]{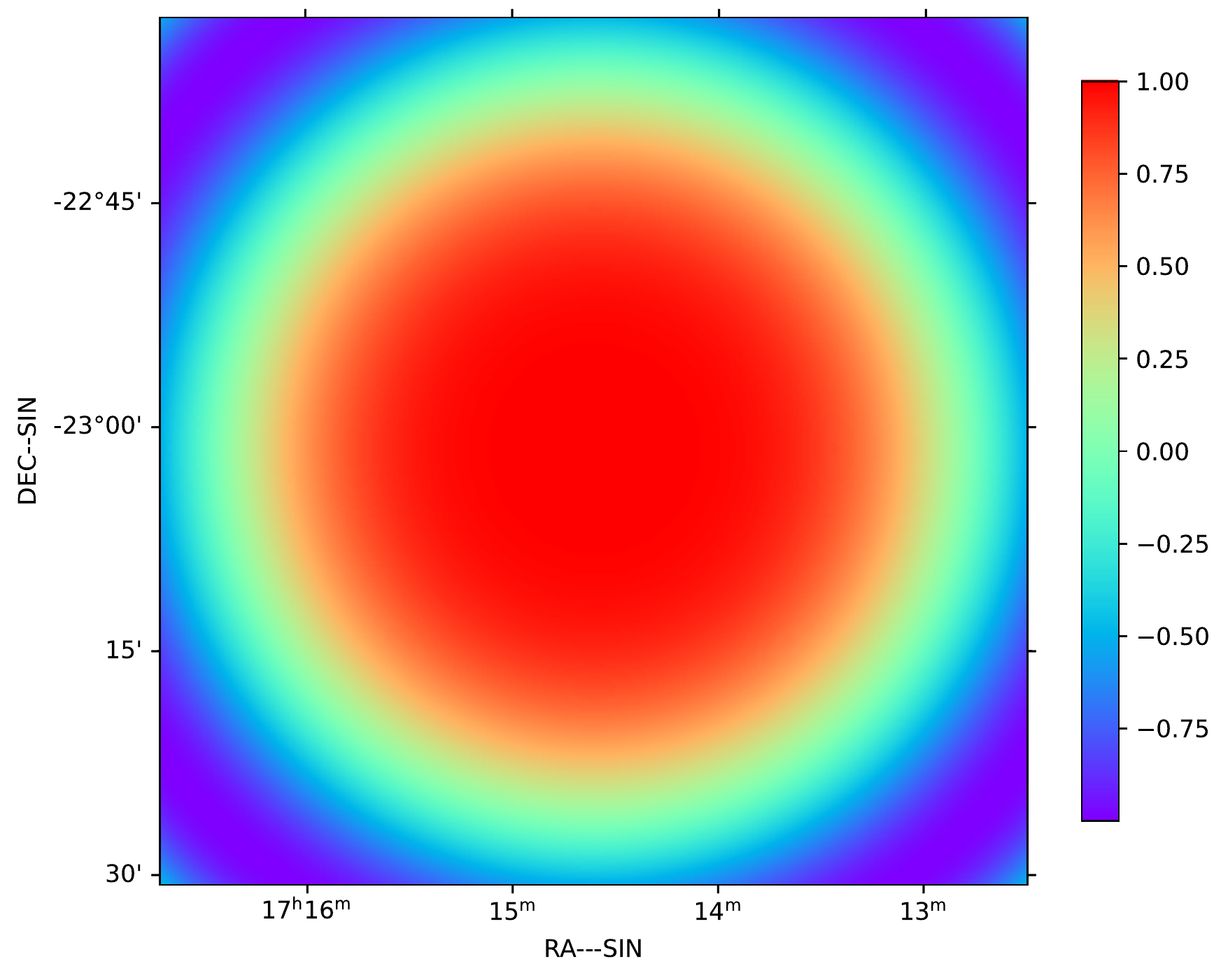}
  \end{minipage}
  \caption{ Diagrams of phase shifts with different baselines and phase screen of maximum $w$. (a) shows the value of phase shifts. (b) shows the phase screen at 12:00, December 11, 2020.}
  \label{fig:phaseshift}
\end{figure}

The non-coplanar baseline effects will directly affect the phase of the visibility data at different distances from the center of the solar disk. 
We analyzed the phase shifts of MUSER using Equation~\ref{wterm} and two extremes of the w-value already obtained, i.e., the maximum value of 9412.12 wavelength numbers and the minimum value of 3156.74 wavelength numbers. The observational frequency was 1GHz as well.
In Figure\ref{fig:phaseshift}(a), the horizontal axis represents the field of view. Moreover, the midpoint of the horizontal axis is the phase center, and the value of the vertical axis represents the value of phase shift. It can be seen that at the center of FOV, the phase shift is 0. However, as the distance from the phase center increases, the phase shift will keep increasing. And the larger the value of $w$, the larger the phase shift increases with the distance. We could find that the phase shift of the maximum $w$ value(9412.12) is almost three times that of the minimum $w$ value(3156.74) at the FOV of 32'. 
Furthermore, it can be derived that at the location of the solar radius (16'), the phase shift increases by 0.2785 radians with respect to the location of 12', while at the location of 32', the phase shift increases by 1.9215 radians compared to the location of 16'.

To make the phase variation more intuitive, we give the phase screen image ($e^{2\pi iw(\sqrt{1-l^2-m^2}-1)}$)for December 11, 2020 at 12:00:00 (CST), with $w$ value is 9412.12, as shown in Figure~\ref{fig:phaseshift}(b). When the value phase shift is 0, the image value is 1. 

\section{Imaging Quality Assessment}
\label{sect:imaging}
To analyze the impact of different w-term values on the final imaging and to provide a reasonable basis for subsequent MUSER data processing, we adopted a data simulation approach using the Radio Astronomy Simulation, Calibration and Imaging Library (RASCIL\footnote{\url{https://gitlab.com/ska-telescope/external /rascil}}) to simulate the MUSER-I imaging observations. RASCIL is a fully open-source radio interferometric array data processing package that has been widely used in SKA data simulation and processing studies. Its processing results have been compared with other software and confirmed to be reliable. The RASCIL is developed in Python. The Gridder uses the corresponding module of Ducc\footnote{\url{https://gitlab.mpcdf.mpg.de/mtr/ducc}}.

\begin{figure*}[htbp]
    \begin{minipage}[]{0.495\linewidth}
        \centering
        \includegraphics[width=80mm]{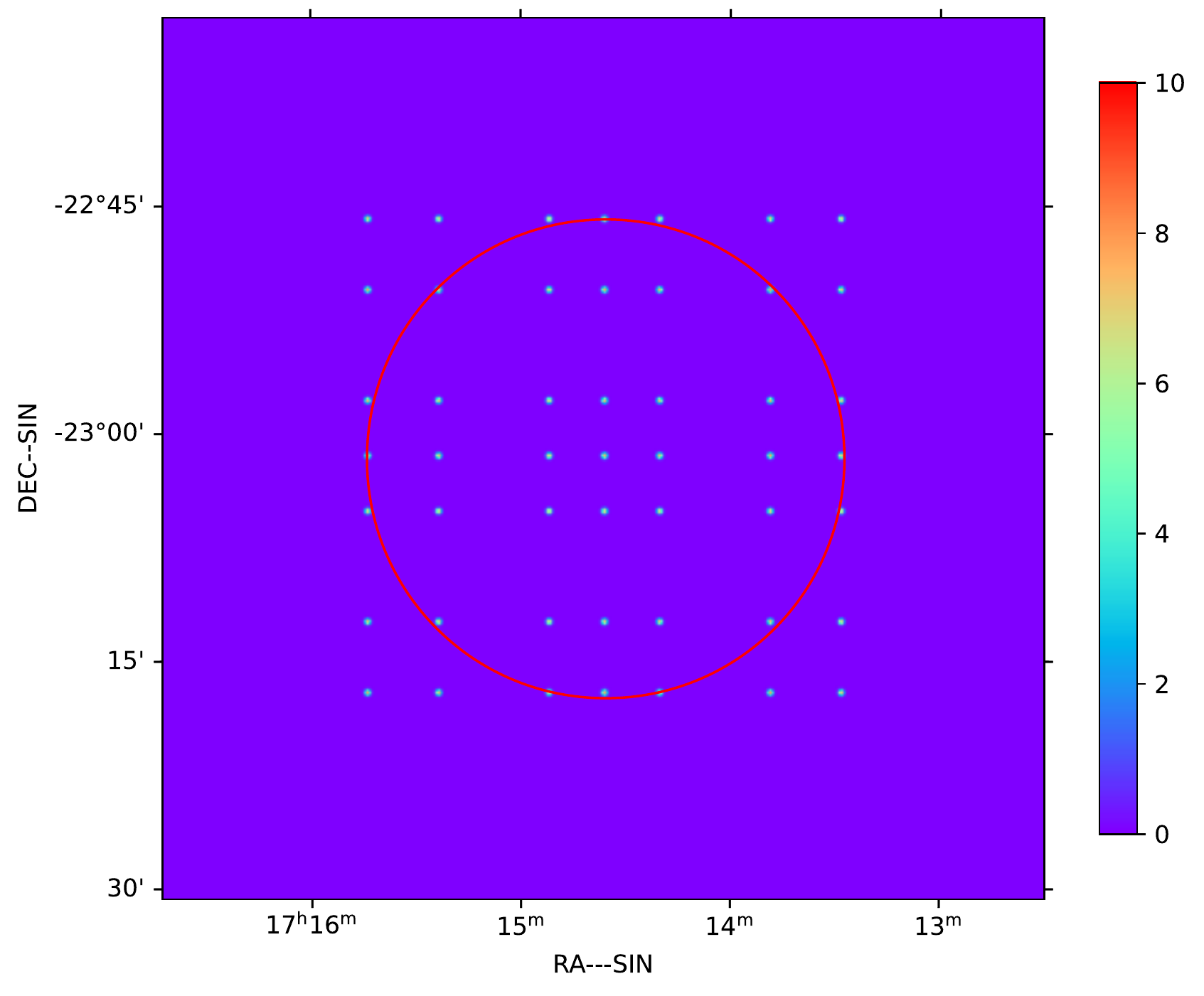}
    \end{minipage}
    \begin{minipage}[]{0.495\linewidth}
        \centering
        \includegraphics[width=65mm]{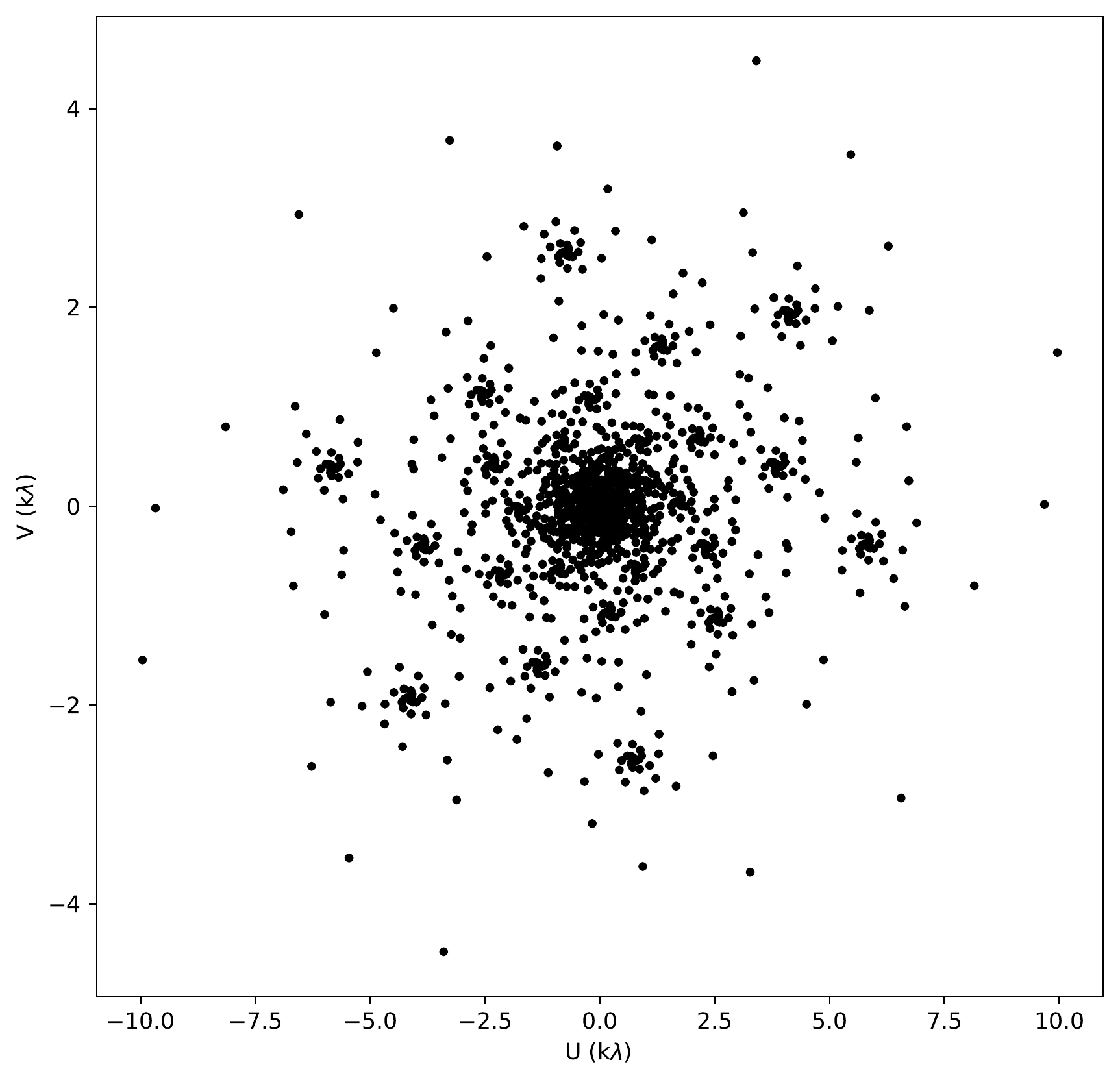}
    \end{minipage}
    \caption{(a) sky model of 2020-12-11-12:00:00(CST).(b)UV coverage of MUSER-I . }
    \label{fig:3mimage}
\end{figure*}

We defined a MUSER-I configuration file for RASCIL, giving the longitude, latitude, and altitude values of MUSER-I. In addition, we carefully configured the antenna diameter, polarization mode, coordinates, and other related parameters.
In the simulated observations, the observation frequency is 1 GHz, and the polarization mode is set to Stocks-I. The sky model used for the simulation consists of 49 point sources (see Figure~\ref{fig:3mimage} (a)), and the flux of point sources is 10 Jy/beam. The distance between the leftmost and rightmost point sources is about 32', which is consistent with the solar diameter. Based on the results of the previous section, we selected 12:00 (CST) on December 11, 2020, for the simulation observations as well. The right ascension of observation is 258.6503(degrees), and the declination of observation is -23.0267(degrees). Meanwhile, the observation hour angle is 6.2296(radians). The observation uv coverage is in Figure\ref{fig:3mimage}(b). The imaging cell size is 6.81 arcsecond, the image size is $512\times512$. Therefore the FOV of the image is 0.968 degrees. 

During the processing of simulated observational data, at first we adopted the classical imaging method of radio interferometry (see Equation\ref{viscal2d}), i.e., gridding the received visibilities, and then obtaining the dirty image by Fourier transform. The non-coplanar baseline effects are not corrected in this imaging process. Then the w-projection~\citep{Cornwell2008, Muscat2014} algorithm is used to imaging, meanwhile to correct the non-coplanar baseline effects. In w-projection, the number of $w$-layers is obtained by using the methods of \cite{Humphreys2011}.

\begin{figure*}[htbp]
    \begin{minipage}[]{0.495\linewidth}
        \centering
        \includegraphics[width=80mm]{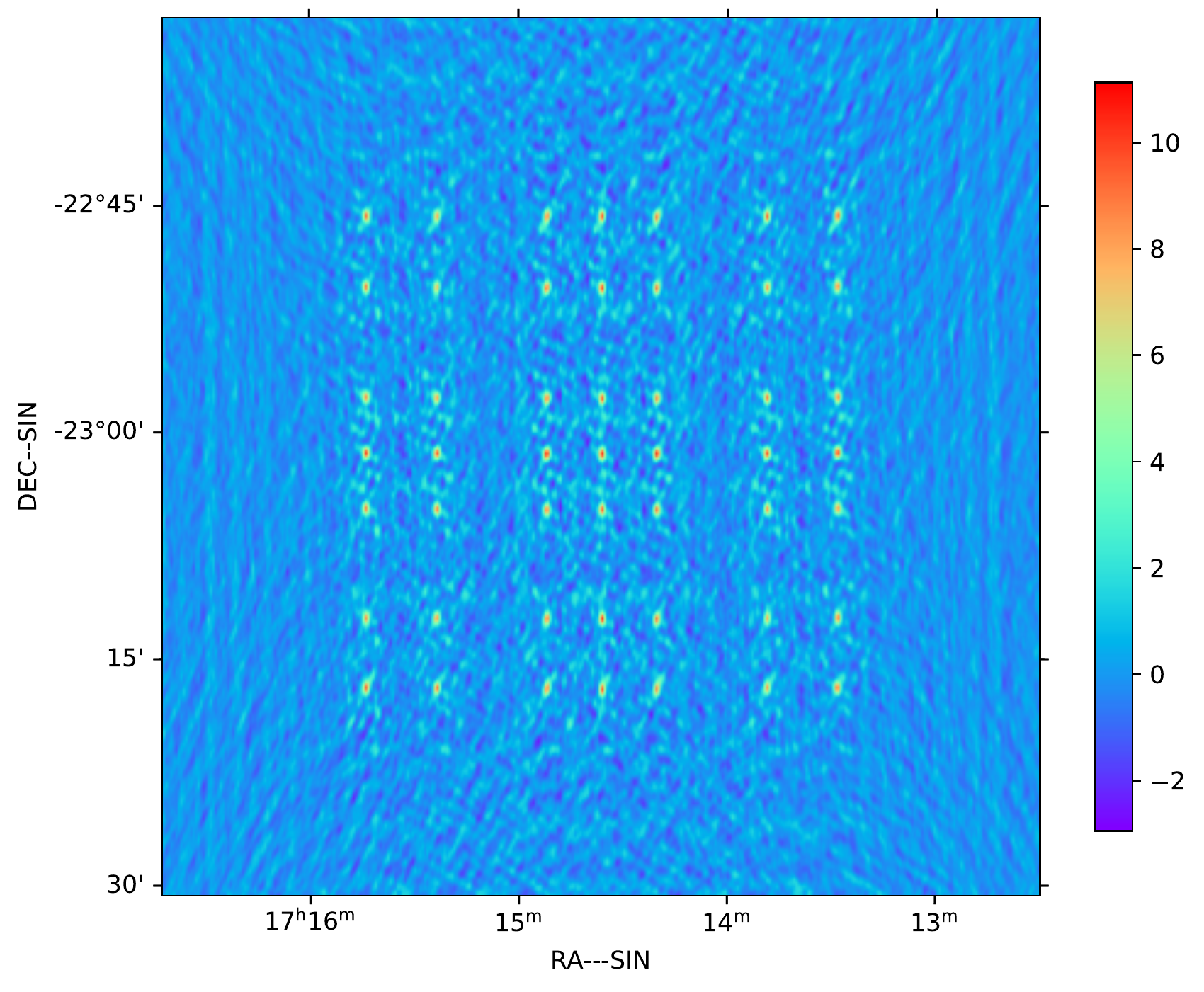}
    \end{minipage}
    \begin{minipage}[]{0.495\linewidth}
        \centering
        \includegraphics[width=80mm]{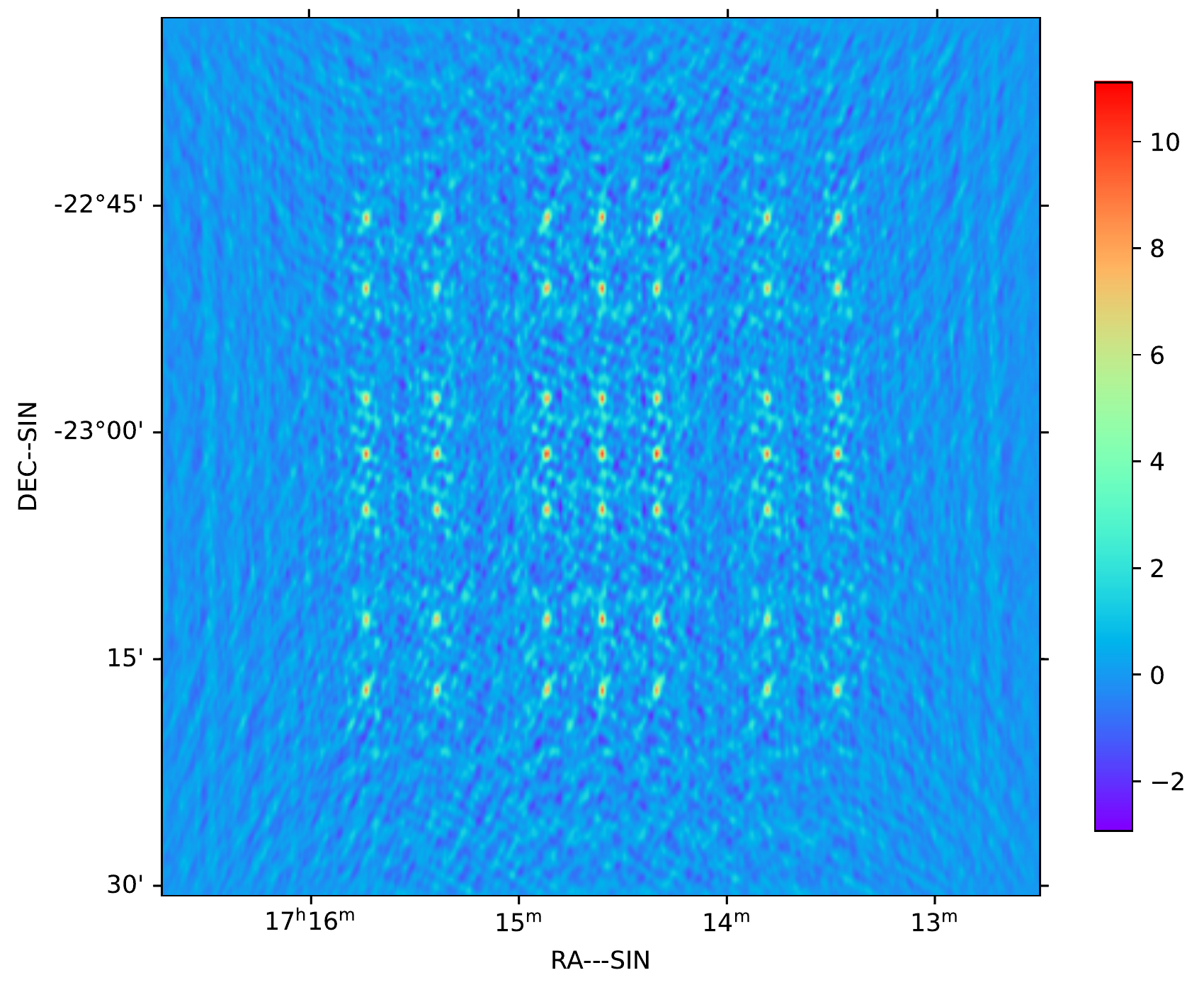}
    \end{minipage}
    \begin{minipage}[]{0.495\linewidth}
        \centering
        \includegraphics[width=80mm]{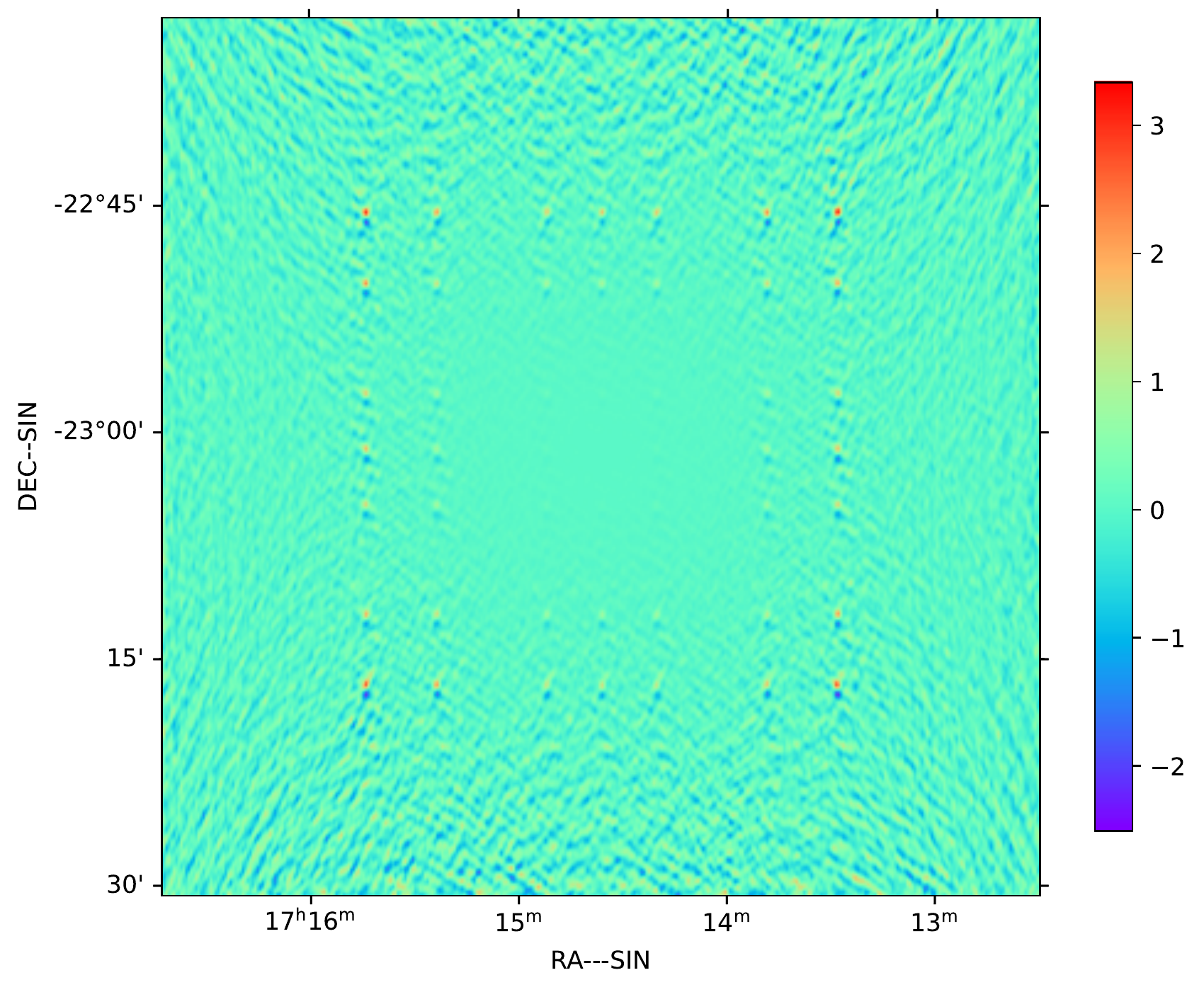}
    \end{minipage}
    \begin{minipage}[]{0.495\linewidth}
        \centering
        \includegraphics[scale=0.37,keepaspectratio=true]{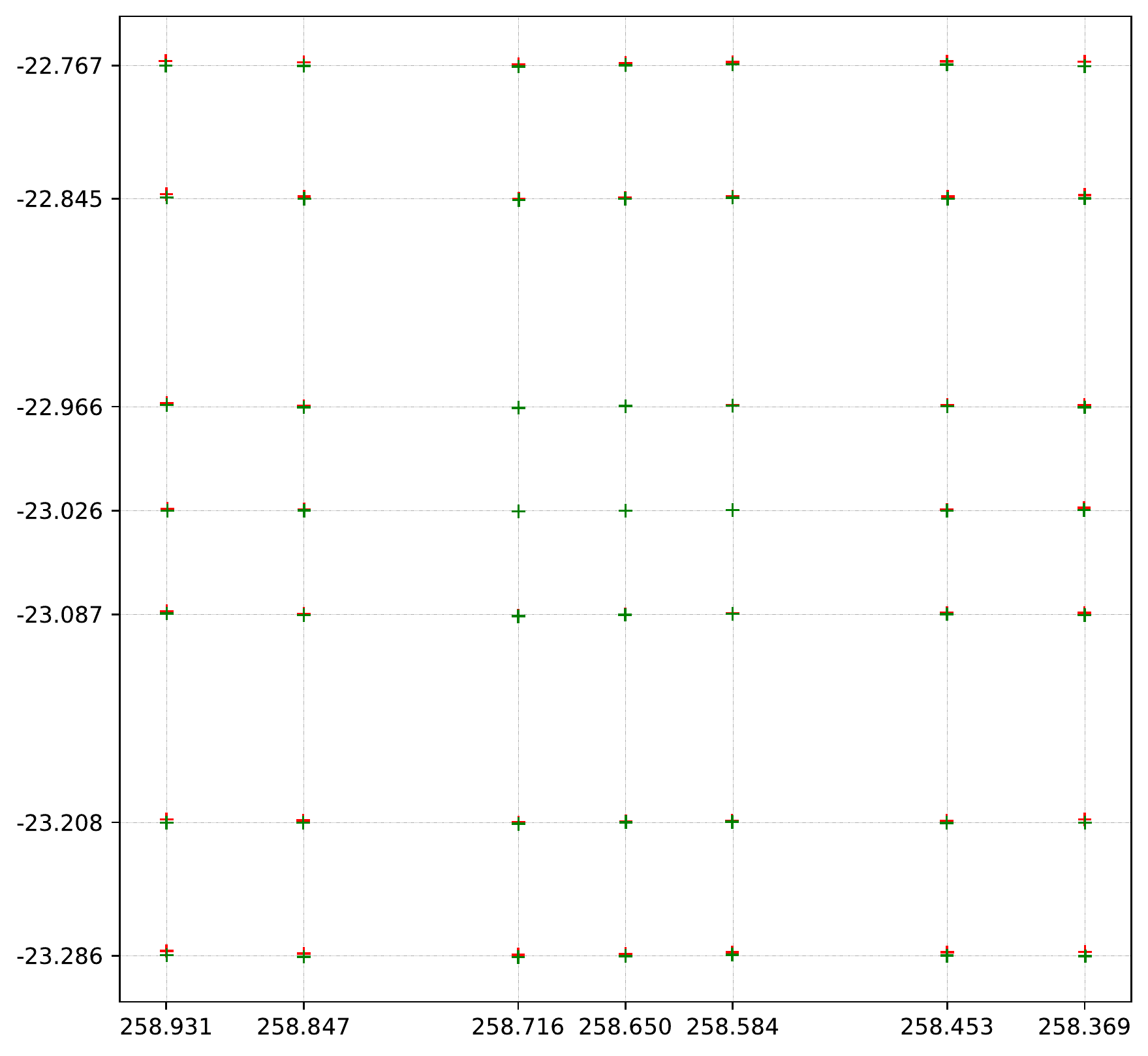}
    \end{minipage}
    \caption{(a) dirty image with non-coplanar baseline effects.(b)dirty image without non-coplanar baseline effects(w-projection). (c)the differences image of dirty image with/without non-coplanar baseline effects.(d) the sources centroid locations in dirty images.}
    \label{fig:dirtyim}
\end{figure*}
After the simulation observation and imaging processing, we present the corresponding results in Figure~\ref{fig:dirtyim}. Figure~\ref{fig:dirtyim}(a) and Figure~\ref{fig:dirtyim}(b) shows the dirty images with/without non-coplanar baseline effects, respectively. Figure~\ref{fig:dirtyim}(c) shows the result of subtracting Figure~\ref{fig:dirtyim}(a) and Figure~\ref{fig:dirtyim}(b).

\begin{figure*}[htbp]
    \centering
   \begin{minipage}[]{0.495\linewidth}
        \centering
        \includegraphics[width=68mm]{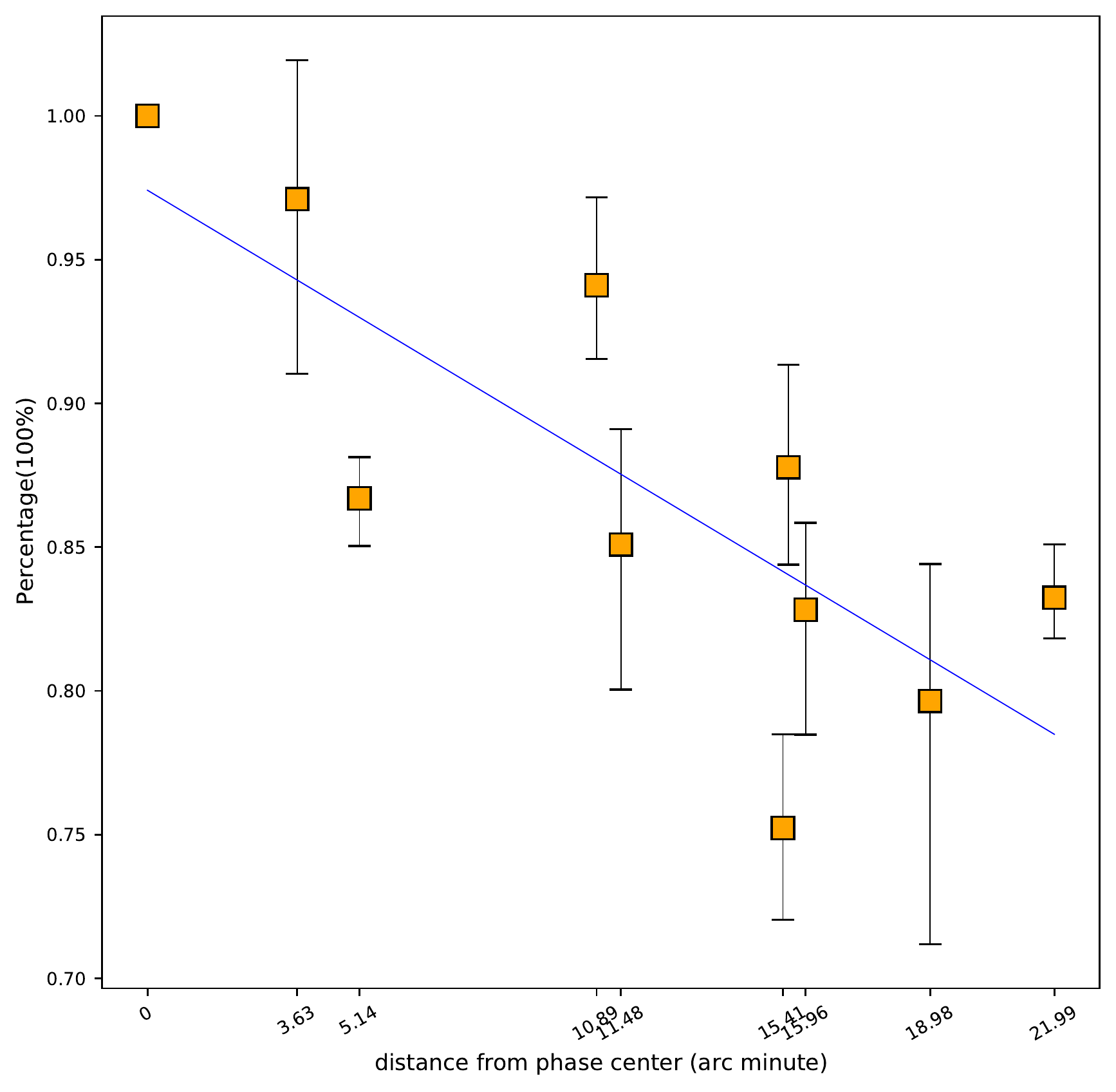}
    \end{minipage}
    \begin{minipage}[]{0.495\linewidth}
        \centering
        \includegraphics[width=68mm]{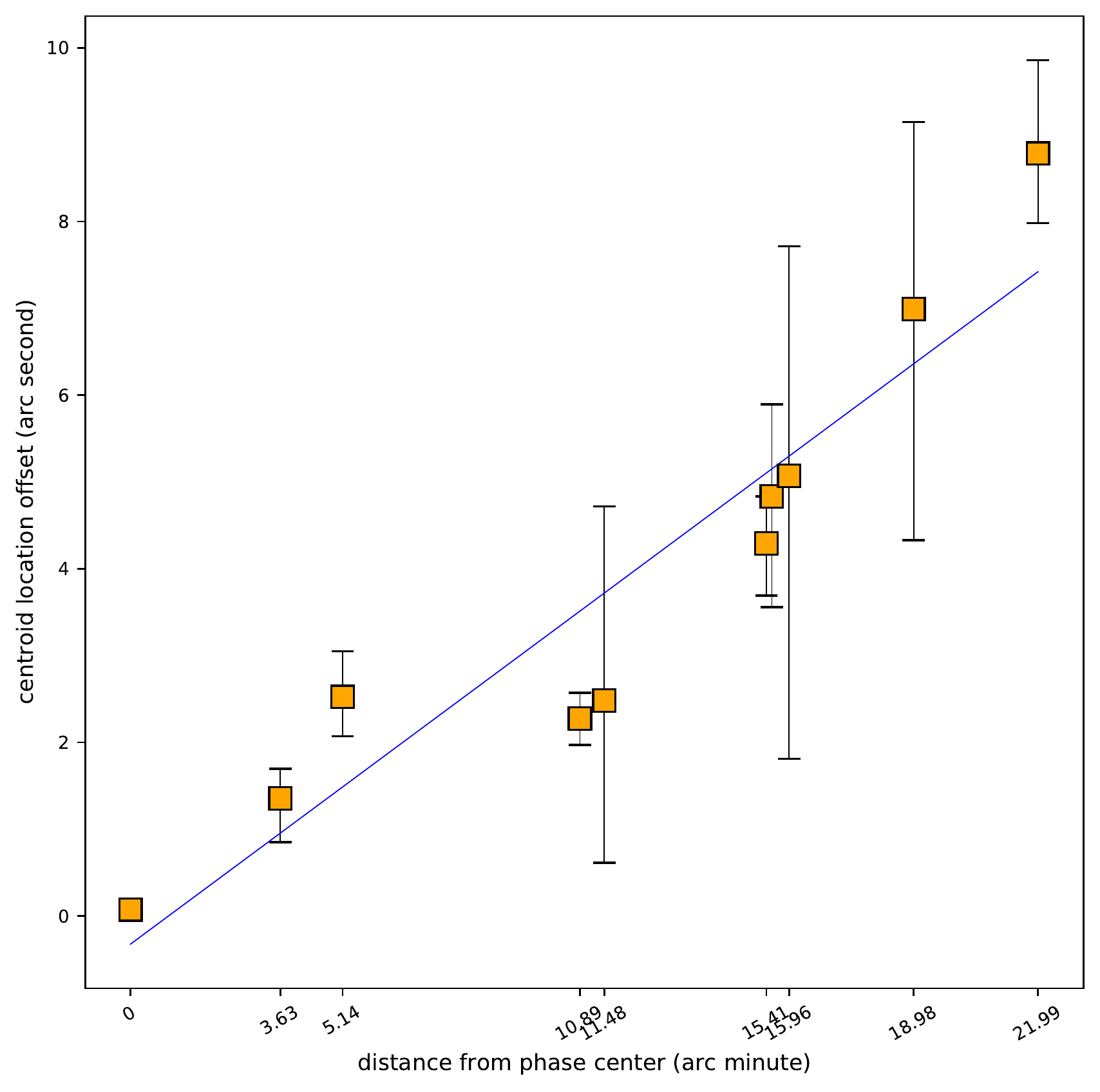}
    \end{minipage}
    \caption{(a) diagram of point source flux attenuation without w-term correction.(b)statistical result of centroid location offset in dirty image without w-term correction.}
    \label{fig:statistical} 
\end{figure*}

In dirty images obtained by both methods, the point source at the phase center  (right ascension:258.6504, declination:-23.0264, unit:degrees) has a flux of 10.93 Jy/beam, but the point source at the location (258.6504, -22.7673) has a flux of 10.43 Jy/beam in Figure\ref{fig:dirtyim}(a) and 10.81 Jy/beam in Figure\ref{fig:dirtyim}(b). 
Point sources farther from the phase center, such as at the location (258.9313, -22.7673) have a flux of 8.95 Jy/beam in Figure\ref{fig:dirtyim}(a) and 9.82 Jy/beam in Figure\ref{fig:dirtyim}(b). It can be noticed that non-coplanar baseline effects would have different effects on fluxes of point sources at different locations. Therefore, in the dirty image without w term correction, we further investigated the relationship between the fluxes of 49 point sources and their distances from the phase center. In Figure\ref{fig:statistical}(a) 
the horizontal axis denotes the distances from the locations of the point sources to the phase center, and the vertical axis indicates that the flux attenuation of the point source varies with the distance from the phase center. At a distance of 18.98 arc-minutes from the phase center, the ratio is about 80$\%$.

The point sources at the edge are more severely affected by the non-coplanar baseline effects, and the locations of the point sources are shifted. By calculating the centroid locations of the point sources in the dirty images, we obtained the centroid locations offset distribution of the point sources in both imaging methods. In Figure~\ref{fig:dirtyim}(d), the red markers are the centroid locations of point sources in the dirty image without the non-coplanar baseline effects correction, and the green markers are the centroid locations of point sources in the dirty image, which is obtained by the w-projection algorithm. It can be inferred that the point sources have offset in centroid locations when non-coplanar baseline effects are not corrected in the imaging process.

Therefore, we statistic the centroid locations offset data for all point sources in Figure\ref{fig:dirtyim}(a), and the results are shown in Table\ref{tb:offset}. We could see that the centroid locations offset in right ascension is 0.494 arc seconds, and in declination is 9.848 arc seconds at the location (258.9313, -23.2858). 
Furthermore, Figure\ref{fig:statistical}(b) shows the relationship between the point source location and the mean value of the point source centroid location offset. At a distance 18.98 arc minutes from the phase center, the offset mean value is about 7 arc seconds. It can be concluded that the farther the point sources are from the phase center, the larger the centroid position offsets due to the non-coplanar baseline effect.

\begin{table*}[htbp]
\caption{Offset of point sources centroid locations in the dirty image without non-coplanar baseline effects correction (unit:arc-second)} 
\label{tb:offset}
\resizebox{\linewidth}{!}{
\begin{tabular}{cccccccc}
\hline
RA$\backslash$ DEC(degrees)   & -23.28578874  & -23.20799939  & -23.08696216  & -23.02644368  & -22.9659252  & -22.844888  & -22.76734809\\
\hline
  258.9313419  & (0.494, 9.848)  & (0.413, 6.573)  & (0.851,  6.887)  & (1.990, 4.108)  & (0.846, 7.669)  & (0.257, 9.146)  & (-1.429, 9.196)\\
  258.8472499  & (0.190, 4.327)  & (-1.169, 3.986) & (0.124, 1.583)   & (0.704, 2.468)  & (0.286, 1.136)  & (0.884, 4.402)  & (0.321, 6.373)\\
  258.7159846  & (-0.170, 1.803) & (0.608, -0.088) & (-0.005,-3.049)  & (0.383, -1.652) & (0.462, -2.110) & (1.223, -0.275) & (0.759, 2.826)\\
  258.6503521  & (0.025, 3.560)  & (0.719, 2.469)  & (-0.768, -0.367) & (0.066, -0.030) & (0.323, 1.459)  & (-0.808, 1.815) & (0.075, 5.320)\\
  258.5847195  & (-1.093, 6.825) & (-1.362, 3.791) & (-0.695, 1.950)  & (-0.226, 1.361) & (-0.067, 2.821) & (-0.459, 4.698) & (-0.173, 7.392)\\
  258.4534543  & (-0.513, 6.872) & (-1.224, 3.482) & (-0.790, 3.267)  & (-0.772, 1.814) & (-0.240, 3.119) & (0.907, 4.749)  & (-0.794, 8.811)\\
  258.3693622  & (0.834, 7.954)  & (-0.076, 6.122) & (-0.487, 3,295)  & (-1.322, 5.746) & (-0.398, 3.530) & (-0.008, 7.637) & (-0.330, 7.976)\\
\hline
\end{tabular}
}
\end{table*}

From the above imaging results, it can be concluded that MUSER is affected by the non-coplanar baseline effect during imaging. This effect is more noticeable when the $w$ value is larger. 
Moreover, since the non-coplanar baseline effect is proportional to the FOV size, the edge part of the solar disk is more affected than the central part (phase center).

\section{Conclusions and Future work}
\label{sect:conclusion}

In this study, we investigate the non-coplanar baseline effects on MUSER, which has an almost 3000-meter long baseline. Experiments on simulation observation have shown that the maximum $w$ values of MUSER are from 3156.74 (number of wavelengths) to 9412.12 (number of wavelengths), which would affect the imaging quality of observation. We demonstrate the non-coplanar baseline effects in the imaging process by two imaging algorithms, including the classical Fourier transform algorithm and w-projection. It can be concluded that the non-coplanar baseline effects should be considered in MUSER's imaging processing for almost all the year if we want to get a high dynamic and finely solar image. 

For radioheliograph imaging, a long baseline can improve the imaging resolution. However, this study presents that although the field of view is small for solar observations, non-coplanar effects can significantly impact solar observation imaging, especially for measurements related to precise positions.

Furthermore, to obtain higher quality observation images for MUSER, investigating of imaging algorithms for correcting non-coplanar baseline effects should be taken into account. There are two issues to be considered. The first is to compare and choose proper methods for non-coplanar baseline effects correction. Besides the w-projection mentioned above, there are additional imaging algorithms that are used to correct the non-coplanar baseline effects. As these algorithms have different features, it is necessary to find one that is the best suitable for the MUSER data processing pipeline. The second is to meet the requirements of real-time imaging. MUSER-I would output 1.92 GB of observation data per minute and MUSER-II would reach 3.6 GB per minute \citep{Wang2015}. How to correct the non-coplanar baseline effects for these data in real-time is also a problem of MUSER's data processing pipeline. In the future, the focus will be on the running efficiency of various imaging algorithms, and apparently, parallel computing based on the multi-core CPU or GPU will be a good choice.

\begin{acknowledgements}
This work is supported by the National SKA Program of China (2020SKA0110300), the Joint Research Fund in Astronomy (U1831204, U1931141) under cooperative agreement between the National Natural Science Foundation of China (NSFC) and the Chinese Academy of Sciences (CAS), the Funds for International Cooperation and Exchange of the National Natural Science Foundation of China (11961141001), the National Natural Science Foundation of China (No.11903009).

\end{acknowledgements}

\bibliographystyle{raa}

\begin{thebibliography}{99}

  \bibitem[Ryle (1962)]{Ryle1962} Ryle M.,1962 ,Nature,194,517

  \bibitem[Bastian (2003)]{Bastian2003} Tim S. Bastian, 2003, Innovative Telescopes and Instrumentation for Solar Astrophysics, proceedings of SPIE, Vol.4583, doi:10.1117/12.460293
  
  \bibitem[Bhatnagar (1999)]{Bhatnagar1999} Sanjay Bhatnagar,1999,Notes from summer school on low frequency astronomy at NCRA, Pune
  
  \bibitem[Bhatnagar et al. (2008)]{Bhatnagar2008} S. Bhatnagar , T. J. Cornwell,K. Golap , J. M. Uson,2008, \aap, 487,419
  
  \bibitem[Cornwell(2008)]{Cornwell2008} Tim J. Cornwell, 2008, IEEE Journal of Selected Topics in Signal Processing, 2,793
  
  \bibitem[Cornwell et al. (2005)]{Cornwell2005} Tim J. Cornwell , K. Golap , S. Bhatnagar ,2005,
Astronomical Society of the Pacific Conference Series, 347

  \bibitem[Gary (2003)]{Gary2003} D. Gary, 2003,Journal of The Korean Astronomical Society,36,135

  \bibitem[Grechnev et al. (2003)]{Grechnev2003} V.V. Grechnev , S.V. Lesovoi , G. Ya. Smolkov et al.,2003, Solar Physics,216,239

  \bibitem[Humphreys and Cornwell (2011)]{Humphreys2011} B. Humphreys , T. J. Cornwell,2011,Square Kilometre Array Memo,132

  \bibitem[Kerdraon and Delouis (1997)]{Kerdraon1997} Kerdraon Alain , Delouis Jean-Marc,1997,
The Nan{\c{c}}ay Radioheliograph, Coronal physics from radio and space observations, 192

  \bibitem[Kerdraon and Klein (2011)]{Kerdraon2011} Kerdraon Alain , Klein Ludwig Karl,2011, General Assembly and Scientific Symposium, 2011 XXXth URSI

  \bibitem[Mei et al. (2017)]{Mei2017} Mei Ying , Wang Feng , Wang Wei  , Chen Linjie et al.,2017,\pasp, 130,014503

  \bibitem[Muscat (2014)]{Muscat2014} Muscat Daniel,2014, arXiv preprint arXiv:1403.4209

  \bibitem[Nakajima et al. (1994)]{Nakajima1994} H. Nakajima , M. Nishio , S. Enome, K. Shibasaki , T. Takano , Y. Hanaoka  et al.,1994, Proc.IEEE,82,702

  \bibitem[Offringa et al. (2014)]{Offringa2014} A.R. Offringa , B. McKiley et al.,2014,MNRAS, 444,606

  \bibitem[Perley (1989)]{Perley1989} R. A. Perley,1989, Astronomical Society of the Pacific Conference Series,6

  \bibitem[Perley (1999)]{Perley1999} R. A. Perley, 1999,Synthesis imaging in radio astronomy II,180,383

  \bibitem[Ramesh et al. (1998)]{Ramesh1998} R. Ramesh , K. R. Subramanian ,  M. S. Sundararajan , C. V. Sastry,1998, Sol.Phys,181,439

  \bibitem[Sault (1999)]{Sault1999} R. J. Sault et al., 1999, \aap, 139,387

  \bibitem[Taylor et al. (1999)]{taylor1999synthesis} Greg B.Taylor ,  Chris Luke Carilli, Richard A Perley,1999, Synthesis Imaging in Radio Astronomy II,180

  \bibitem[Thompson et al. (2008)]{Thompson2008}  A. R. Thompson,  J. M. Moran, G.W.Swenson Jr, 2008, John Wiley \& Sons

  \bibitem[Wang et al. (2015)]{Wang2015} Wang Feng, Mei Ying, Wang Wei , Ji, Kaifan,2015,\pasp,127,950

  \bibitem[Wang and Yan (2019)]{Wang2019}  Wang Wei, Yan Yihua, 2019,Advances in Astronomy,2019

  \bibitem[Wei et al. (2016)]{Wei2016} Wei Shoulin , Wang Feng , Deng Hui,Liu Cuiyin , DaiWei, Liang Bo ,Mei Ying,Shi Congming, Liu Yingbo ,Wu Jingping,2016,\pasp,129,024001

  \bibitem[Yan et al. (2015)]{Yan2015} Yan Yihua ,Chen Linjie,Yu Sijie,2015,International Astronomical Union. Proceedings of the International Astronomical Union,11,427

  \bibitem[Yan et al. (2011)]{Yan2011} Yan Yihua ,Zhang Jian , Chen Zhijun , Wang Wei, Liu Fei, Geng Lihong, 2011,General Assembly and Scientific Symposium, 2011 XXXth URSI

  \bibitem[Yan et al. (2004)]{Yan2004} Yan Yihua, Zhang Jian,Huang Guangli,2004, Radio Science Conference, 2004. Proceedings. 2004 Asia-Pacific

  \bibitem[Yan et al. (2021)]{Yan2021} Yan Yihua ,Chen Zhijun,Wang Wei,Liu Fei,  Geng Lihong,Chen Linjie , Tan Chengming ,Chen Xingyao et al., 2021,Frontiers in Astronomy and Space Sciences

  

\end{thebibliography}

\label{lastpage}

\end{document}